\documentclass[pre,twocolumn,superscriptaddress,floatfix,showpacs,showkeys,nofootinbib]{revtex4}

\usepackage[latin2]{inputenc}
\usepackage{color}
\usepackage{amsmath}
\usepackage{amssymb}
\usepackage[dvips]{graphicx}
\usepackage{epsfig}
\usepackage{mathptmx}      
\usepackage{float}
\usepackage{tabularx}
\usepackage{array}

\newcommand{\muf}{$\mu_\text{eff}$}

\newcommand{\Muf}{\mu_\text{eff}}
\newcommand{\Mup}{\mu_\text{P}}
\newcommand{\Muw}{\mu_\text{W}}

\newcommand{\vcone}{V_{BC}}
\newcommand{\vctwo}{V_{AB}}

\begin{document}

\title{Shear flow of dense granular materials near smooth walls. I.\\Shear
    localization and constitutive laws in boundary region}

\date{\today}

\author{Zahra Shojaaee}
\email{zahra.shojaaee@uni-duisburg-essen.de}
\affiliation{Faculty of Physics, University of Duisburg-Essen, 47048 Duisburg, Germany}

\author{Jean-No\"el Roux}
\affiliation{Universit\'e Paris-Est, Laboratoire Navier (IFSTTAR, Ecole des Ponts ParisTech, CNRS), 2 All\'ee Kepler, 77420 Champs-sur-Marne, France}

\author{Fran\c{c}ois Chevoir}
\affiliation{Universit\'e Paris-Est, Laboratoire Navier (IFSTTAR, Ecole des Ponts ParisTech, CNRS), 2 All\'ee Kepler, 77420 Champs-sur-Marne, France}

\author{Dietrich E. Wolf}
\affiliation{Faculty of Physics, University of Duisburg-Essen, 47048 Duisburg, Germany}

\begin{abstract}
  We report on a numerical study of the shear flow of a simple
  two-dimensional model of a granular material under controlled normal
  stress between two parallel smooth, frictional walls, moving with
  opposite velocities $\pm V$. Discrete simulations, which are carried
  out with the contact dynamics method in dense assemblies of disks,
  reveal that, unlike rough walls made of strands of particles, smooth
  ones can lead to shear strain localization in the boundary
  layer. Specifically, we observe, for decreasing $V$, first a
  fluid-like regime (A), in which the whole granular layer is sheared,
  with a homogeneous strain rate except near the walls; then (B) a
  symmetric velocity profile with a solid block in the middle and
  strain localized near the walls and finally (C) a state with broken
  symmetry in which the shear rate is confined to one boundary layer,
  while the bulk of the material moves together with the opposite
  wall. Both transitions are independent of system size and occur for
  specific values of $V$. Transient times are discussed. We show that
  the first transition, between regimes A and B, can be deduced from
  constitutive laws identified for the bulk material and the boundary
  layer, while the second one could be associated with an instability
  in the behavior of the boundary layer. The boundary zone
  constitutive law, however, is observed to depend on the state of the
  bulk material nearby.
\end{abstract}



\maketitle

\section{Introduction}
\label{sec-intro}

An active field of research over the last three
decades~\cite{Campbell_2006,Forterre_08}, the rheology of dense
granular flows recently benefitted from the introduction of robust and
efficient constitutive laws. First identified in plane homogeneous
shear flow~\cite{daCruz_etal_05}, those laws were successfully applied
to various flow geometries~\cite{Midi_04}, such as inclined
planes~\cite{Forterre_08}, or annular shear
devices~\cite{Koval_etal_09}, both in numerical and experimental
works~\cite{Lacaze_09}. A crucial step in the formulation of these
laws is the characterization of the internal state of the
homogeneously sheared material in steady flow under given normal
stress by the \emph{inertial number}~$I$~\cite{Midi_04,daCruz_etal_05}
(see also Eq.~(\ref{inertial_number})), expressing the ratio of shear
time to rearrangement time, thereby regarding the material state as a
generalization of the quasistatic critical state, which corresponds to
the limit of $I\to 0$. Once identified in one geometry, those
constitutive laws prove able to predict velocity fields and various
flow behaviors in other situations, with no adjustable
parameter~\cite{Pouliquen_etal_06}.

However, assuming a general bulk constitutive law to be available, in
general, one needs to supplement it with suitable boundary conditions
in order to solve for velocity and stress fields in given flow
conditions. Recent studies, mostly addressing bulk behavior, tended to
use rough boundary surfaces, both in experiments (as
in~\cite{Savage_Sayed_84, Hanes_Inman_85, Pouliquen_99}) and in
simulations~\cite{Silbert_etal_01, Lois_etal_05, daCruz_etal_05,
  Taberlet_etal_08, Koval_etal_09}, in order to induce deformation
within the bulk material and study its rheology. Yet, in practical
cases, such as hopper discharge flow~\cite{Nedderman_92}, granular
materials can be in contact with smooth walls (i.e., with asperities
much smaller than the particle diameter), in which case some slip
(tangential velocity jump) is observed at the
wall~\cite{Artoni_etal_09, Zheng_Hill_98, Louge_94,
  Shojaaee_etal_11-2}, and the velocity components parallel to the
wall can vary very quickly over a few grain diameters. The specific
behavior of the layer adjacent to the wall should then be suitably
characterized in terms of a boundary zone constitutive law in order to
be able to predict the velocity and stress fields.

In this work we use grain-level discrete numerical simulation to
investigate the behavior of a model granular material in plane shear
between smooth parallel walls, a simple setup which has already been
observed to produce~\cite{Shojaaee_07, Shojaaee_etal_07,
  Shojaaee_etal_09}, depending on the control parameters, several
possible flow patterns, with either bulk shear flow, or localization
of gradients at one or both walls. We extract a boundary layer
constitutive law similar to the one applying to the bulk material. The
stability of homogeneous shear profiles and the onset of localized
flows at one or both opposite walls have been also
investigated. Although Couette flow between parallel flat smooth walls
is not an experimentally available configuration, we find it
convenient as a numerical test apt to probe both bulk and boundary
layer rheology, and their combined effects on velocity fields and
shear localization patterns.

The structure of the paper is as follows: Sec.~\ref{Sec-System-Setup}
describes the model system that is simulated, and gives the
definitions and methods used to identify and measure various physical
quantities. In Sec.~\ref{section:plane shear flow with smooth walls}
different flow regimes are described, according to whether and how the
velocity gradient is localized near the walls. In
Sec.~\ref{sec-constitutive laws} we derive the constitutive laws both
in the bulk and in the boundary layer. Sec.~\ref{sec:Applications}
applies the constitutive laws identified in Sec.~\ref{sec-constitutive
  laws} to explain some of the observations of Sec.~\ref{section:plane
  shear flow with smooth walls}, such as the occurrence of
localization transitions or the characteristic times associated with
the establishment of steady velocity profiles.
Sec.~\ref{sec-Conclusion-and-Discussion} is a brief conclusion.

\section{System setup}
\label{Sec-System-Setup}

\subsection{Sample, boundary conditions, control parameters}

\begin{figure}[htb]
  \centering
  \includegraphics[width=0.6\columnwidth, clip]{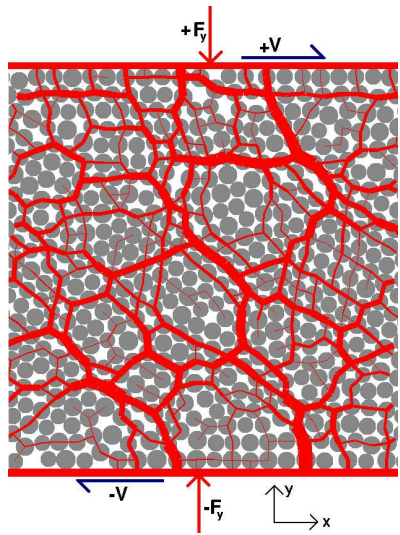}
  \caption{(color online) A polydisperse system of hard frictional
    disks in planar shear geometry with periodic boundary conditions
    in $x$ direction. A prescribed normal force $F_y$ to the confining
    walls, determines the constant external pressure of the
    system. The walls move with the same constant velocity $V$ in
    opposite directions. The width of the lines connecting the centers
    of contacting particles represents the magnitude of the normal
    contact forces above a threshold.}
  \label{fig:System-setup}
\end{figure}

In the contact dynamics method~\cite{Moreau_88,Jean99,
  Brendel_etal_04, Radjai_Richefeu_09} (CD), grains are regarded as
perfectly rigid, and the mechanical parameters ruling contact behavior
are friction and restitution coefficients. The CD method can deal with
dense as well as dilute granular assemblies, and successfully copes
with collisions as well as enduring contacts, and with the formation
and dissociation of clusters of contacting objects. We consider here a
dense assembly of disks (in 2D), with interparticle friction
coefficient $\Mup{=}0.5$. As for dense frictional assemblies, the
restitution coefficient does not influence the constitutive laws
\cite{daCruz_etal_05}, perfectly inelastic collisions are considered
(both normal and tangential restitution coefficients are set to zero
($e_n{=}e_t{=}0$)). With the same contact properties at smooth walls
($\mu_W{=}0.5,e_n{=}e_t{=}0$), slip velocities of the same order of
magnitude as the shear velocity occur. To avoid ordering phenomena,
disks are polydisperse, with diameters uniformly distributed between
$0.8 d$ and $d$. The largest diameter $d$ is taken as the length unit
throughout the following ($d{=}1[L]$). Similarly the mass density of
the particles is set to unity ($\rho {=}1[M]/[L]^2$), so that the mass
of a disk with unit diameter is $m{=}\pi/4$. The time unit is chosen
such that the pressure (normal forces applied to the walls divided by
the length of the walls) have a value
$\sigma_{yy}{=}F_y/L_x{=}0.25[M]/[T]^2$, which leads to:
$F_y{=}5[M][L]/[T]^2$. In other words, we use the following base units
for length, mass and time:
\begin{align}
  [L] &= d, \notag \\
  [M] &= d^2 \ \rho, \notag \\
  [T] &= \sqrt{5\,d^3\,\rho/F_y}. \notag
\end{align}
We consider simple shear flow within rectangular cells with periodic
boundary conditions in the flow direction (parallel to the $x$ axis in
Fig.~\ref{fig:System-setup}). Gravity is absent throughout all our
simulations. The top and bottom walls bounding the cell are
geometrically smooth, but their contacts with the grains are
frictional, with a friction coefficient $\Muw$ set to $0.5$. They move
with constant and opposite velocities ($\pm V$) along direction
$x$. They are both subjected to inwards oriented constant forces $F_y$
normal to their surface, so that in steady state a constant normal
stress $\sigma_{yy}$ is transmitted to the sample. The wall motion in
the normal direction is ruled by Newton's law, involving the wall
mass, equal to $50$, thereby causing the system height $L_y$ to vary
in time. In steady state shear flow, $L_y$ fluctuates about its
average value.

Results from different samples of various sizes are presented
below. System sizes and simulation parameters are listed in
Tab.~\ref{tab:LxLy}.

\begin{table}[t]
  \centering
\begin{tabular}{ m{5mm} m{8mm} m{6mm} m{6mm} m{10mm} m{15mm} m{9mm} m{16mm}}
  \hline \hline
  Idx & $n$  & $L_y$ & $L_x$ & $\sigma_{yy}$ & $V$ & $T_\text{SS}$ &
  $T_\text{Sim}$\\
\hline
1&  511  & 20 & 20 & 0.25 & 0.005-5.00 & 620 & 20000\\
2&  1023  & 40 & 20 & 0.25 & 0.03-30.00 & 2500 & 10000\\
3&  1023  & 40 & 20 & 0.0625 & 0.03-30.00 & 9900 & 10000\\
4&  3199  & 50 & 50 & 0.25 & 0.01-30.00 & 4000 & 8000\\
5&  2047  & 80 & 20 & 0.25 & 0.01-20.00 & 10000 & 4000-12000\\
6&  3071   & 120 & 20 & 0.25 & 0.01-35.00 & 22000 & 6000\\
7&  5119   & 200 & 20 & 0.25 & 0.01-30.00 & 64000 & 13000\\
  \hline \hline
\end{tabular}

\caption{Parameters used in the simulations. $n$ is the number of
  disks in the sample. $T_\text{SS}$ denotes the characteristic time to approach steady
  state according to Eq.~(\ref{eq-T-SS}). $T_\text{Sim}$ is the total
  time simulated in each run.}
\label{tab:LxLy}
\end{table}

As in Refs.~\cite{Midi_04,daCruz_etal_05,Koval_etal_09}, the
dimensionless inertial number, $I$, defined as a reduced form of shear
rate $\dot{\gamma}$:
\begin{gather}\label{inertial_number}
  I=\dot{\gamma}\sqrt{\dfrac{m}{\sigma_{yy}}},
\end{gather}
is used to characterize the state of the granular material in steady
shear flow. In contrast to previous studies \cite{Midi_04,
  daCruz_etal_05}, the shear is not homogeneous in the present case
(because of wall slip and of stronger gradients near the walls), and
in general $\dot{\gamma}$ is different from $\dfrac{2V}{L_y}$. Thus,
the shear rate has to be measured locally. We focus in the present
study on shear localization at smooth walls and try to deduce
constitutive laws in the boundary layer, associating the boundary
layer behavior not only with wall slip, but also with the material
behavior in a layer adjacent to the wall, the internal state of which
might be affected by that of the bulk material.

\subsection{System preparation}
\label{subsec-system-preparation}

To preserve the symmetry of the top and the bottom walls, the system
is horizontally filled. While distance $L_y$ between the walls is kept
fixed, a third, vertical wall is introduced, on which the grains
(which are temporarily rendered frictionless) settle in response to a
``gravity'' force field parallel to the $x$ axis. Then the force field
is switched off, and the free surface of the material is smoothened
and compressed by a piston transmitting $\sigma_{xx}=0.25$ (the same
value as $\sigma_{yy}$ imposed in shear flow), until equilibrium is
approached. System width $L_x$ is determined at this stage. Then the
vertical wall and the piston are removed, the friction coefficients
are attributed their final values $\Mup$ and $\mu_W$ and periodic
boundary conditions in the $x$ direction are enforced. With constant
$L_x$ and variable $L_y$, the shearing starts with velocities $\pm V$
for the walls and an initial linear velocity profile within the
granular layer.

\subsection{Measured quantities}
\label{subsec-Measured-Quantities}

Before presenting the results, we first explain the method used to
measure the effective friction coefficient, the velocity profiles and
the inertial number. For a system in steady state, assuming a uniform
stress tensor in the whole system, a common method to calculate the
effective friction coefficient is to average the total tangential and
normal forces acting on the walls over time and then calculate their
ratio. Another way to calculate the effective friction coefficient is
to consider the components of the stress tensor with its contact,
kinetic and rotational contributions inside the system
\cite{Moreau_97,Koval_etal_09}. The stress in our system is dominated
by contact contributions. Let $\sigma_\text{c}^i$ denote the
total contact stress tensor calculated for each particle $i$ with area
$A_i{=}\pi d_i^2/4$:
\begin{gather}
  \sigma_\text{c}^i=\dfrac{1}{A_i} \sum_{j\neq i}
  \vec{F_{ij}} \otimes\vec{r_{ij}}.
\end{gather}
The summation runs over all particles $j$ having a contact with
particle $i$. $\vec{F_{ij}}$ is the corresponding contact force
and $\vec{r_{ij}}$ denotes the vector pointing from the center of
particle $i$ to its contact point with particle $j$. We used both
methods, but finding no significant difference, we present in all our
corresponding graphs the effective friction coefficient (\muf)
measured in the interior of the system considering all terms of the
stress tensor, although the contact contribution dominates.

Our calculation of the velocity profile accounts for particle
rotations, which contribute to the local velocities averaged in
stripes of thickness $\Delta y{=}1$ along the flow direction, as
follows. To each horizontal stripe centered at $y{=}y'$, we attribute
a velocity by averaging the contributions of all the particles it
contains (partly or completely) \cite{Laetzel_etal_00}:
\begin{gather}\label{v_profile}
  \upsilon_x(y')=\dfrac{\displaystyle\sum\limits_i \int\limits_{S_i}(\upsilon_{ix}+\omega_i
    r_{iy})dS}{\displaystyle\sum\limits_i S_i}.
\end{gather}
$S_i$ denotes the surface fraction of particle $i$ within the stripe,
$\upsilon_{ix}$ its center of mass velocity in $x$ direction,
$\omega_i$ its angular velocity and $r_{iy}$ is the vertical distance
between the center of mass of the particle and a differential stripe
of vertical position $y$ and surface $dS$ within surface $S_i$. The
velocity profiles presented here are also averaged over time intervals
of $\Delta t{=}80$. Those time intervals follow each other directly
without any gap.

In the calculation of the profiles of stress tensor, each particle
contributes to each stripe in proportion to the surface area contained
in the stripe. This corresponds to the scheme used in
\cite{Laetzel_etal_00} and is slightly different from the coarse
graining reviewed in \cite{Goldhirsch_10} in the sense that it is
highly anisotropic (with a coarse graining scale of $L_x\times 1$) and
does not incorporate the (stress free) regions beyond the walls. One
other method is to split the contact contributions proportionally to
their branch vector length within each stripe. One may also cut
through the particles and add up the contact forces of all cut branch
vectors. All three different methods lead to the same results in our
simulations.

\section{Velocity profiles and strain localization}
\label{section:plane shear flow with smooth walls}
\index{plane shear flow with smooth walls|(}

\subsection{Steady state}
\label{subsection:Steady State}
\index{Steady State|(}

\begin{figure}[t]
  \centering
  \includegraphics[width=\columnwidth, clip]{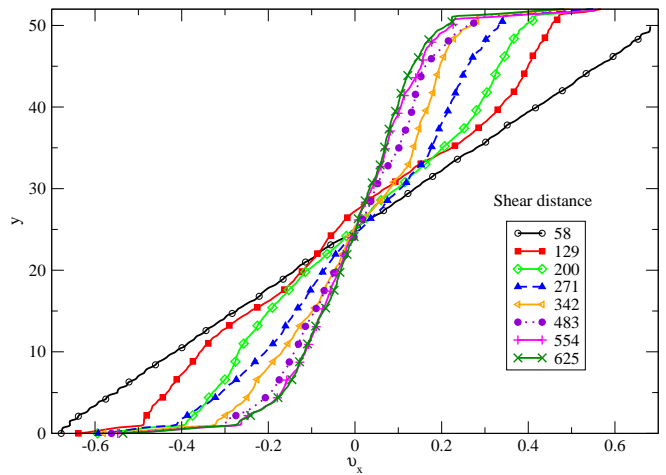}
  \caption{(color online) Transient to steady state for $V=0.70$ in a
    system with $L_x=50$ and $L_y=50$.}
  \label{fig:steady-state-V0.70}
\end{figure}

A system sheared with a certain constant velocity under prescribed
normal stress is expected to reach a steady state after a transient.
For instance, in a system of size $L_x{=}50$ and $L_y{=}50$ with a
large shear velocity, $V{=}0.7$, the steady state is reached after a
shear distance of about $\lambda \simeq 420$, corresponding to a shear
strain of $\gamma \simeq 8$ (Fig.~\ref{fig:steady-state-V0.70}). The
shear distance is calculated by multiplying the total shear velocity
($2V$) by time.  Due to slip at the smooth walls and because of
nonhomogeneous flow, the values attributed to the shear distance and
the shear strain overestimate the real values in the bulk
material. Transient times before steady state will be estimated in
Sec.~\ref{subsec:Transient Time}, based on the constitutive laws.

\begin{figure}[t]
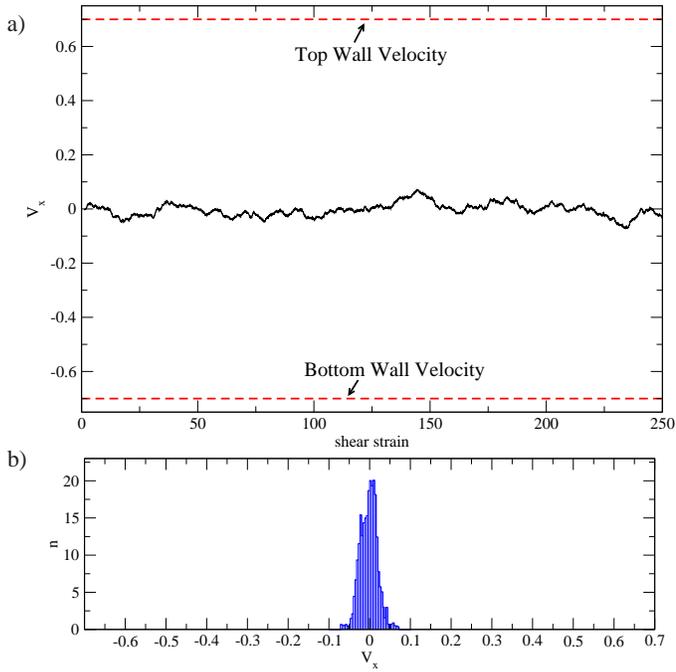

  \centering
  \raisebox{39ex}{a)}\includegraphics[width=\columnwidth, clip]{./Figure-03-a.eps}
  \raisebox{19ex}{b)}\mbox{\hspace{2ex}\includegraphics[width=0.95\columnwidth, clip]{./Figure-03-b.eps}}
  \caption{(color online) (a) Center of mass velocity versus shear
    strain for $V=0.70$ in a system with $L_x{=}50$ and
    $L_y{=}50$. The dashed red lines represent the velocity of the top
    and bottom walls. (b)~Histogram of center of mass velocity
    (accumulated over a long time and over different simulated
    systems).}
    \label{fig:Vx-steady-state-V0.70}
\end{figure}

\begin{figure}[ht]
  \centering
  \includegraphics[width=\columnwidth, clip]{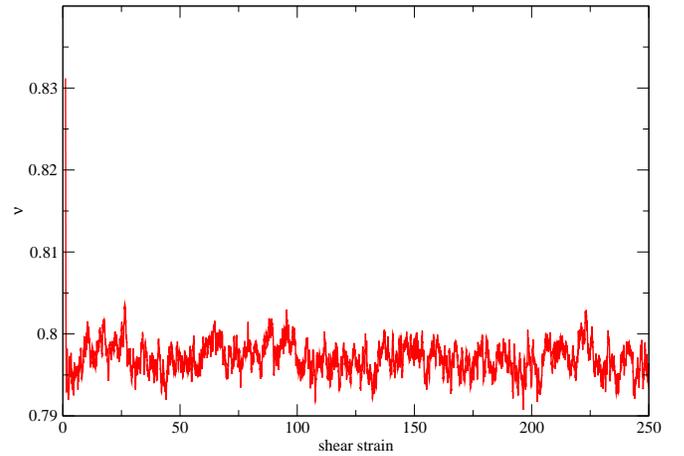}
  \caption{Solid fraction $\nu$ versus shear strain at $V{=}0.70$ in a
    system with $L_x{=}50$ and $L_y{=}50$.}
  \label{fig:Ly-nu-steady-state-V0.70}
\end{figure}

In the steady state, the center of mass velocity in the system of
Fig.~\ref{fig:Vx-steady-state-V0.70} is not conserved, because of the
constant velocity of the walls, and fluctuates about its vanishing
average with an amplitude amounting to about $10\%$ of velocity
$V$. This relatively high level of fluctuation, which is expected to
regress in larger samples, reflects granular agitation within the
sheared layer at large $V$. The fluctuations in height $L_y$ and solid
fraction $\nu$ (measured in the whole system) amount to only about
$1\%$ of the average after a short transient
(Fig.~\ref{fig:Ly-nu-steady-state-V0.70}). The initial sharp drop of
$\nu$, for very small shear strains, is due to the combined effects of
shear flow onset, friction activation and change of boudary conditions
on the configuration prepared as described in
Sec.~\ref{subsec-system-preparation}.

\begin{figure}[t]
  \centering
  \includegraphics[width=\columnwidth, clip]{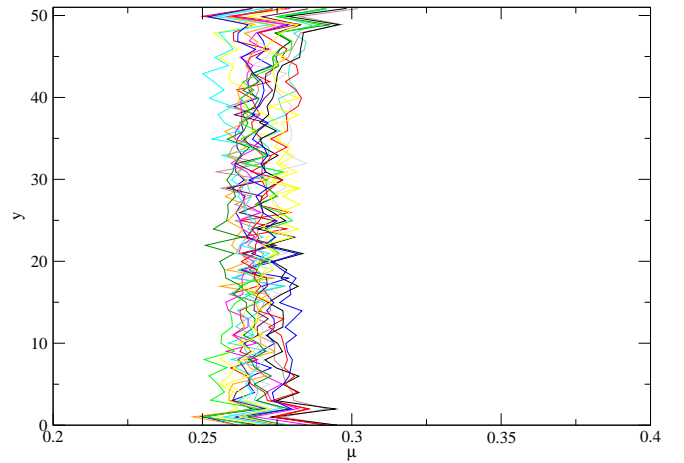}
  \caption{(color online) Profiles of the measured effective friction
    coefficient at different times in steady state for $V{=}0.70$
    ($L_x{=}50$ and $L_y{=}50$).}
  \label{fig:mu-steady-state-V0.70}
\end{figure}

In the steady state the profiles of the effective friction coefficient
stay almost uniform throughout the system
(Fig.~\ref{fig:mu-steady-state-V0.70}), but fluctuate in time, which
is a direct consequent of shearing with constant velocity and
consequent fluctuations in the center of mass velocity.

\subsection{Shear regimes and strain localization}
\label{subsection:Shear Regimes}
\index{Shear Regimes|(}

\begin{figure}[t]
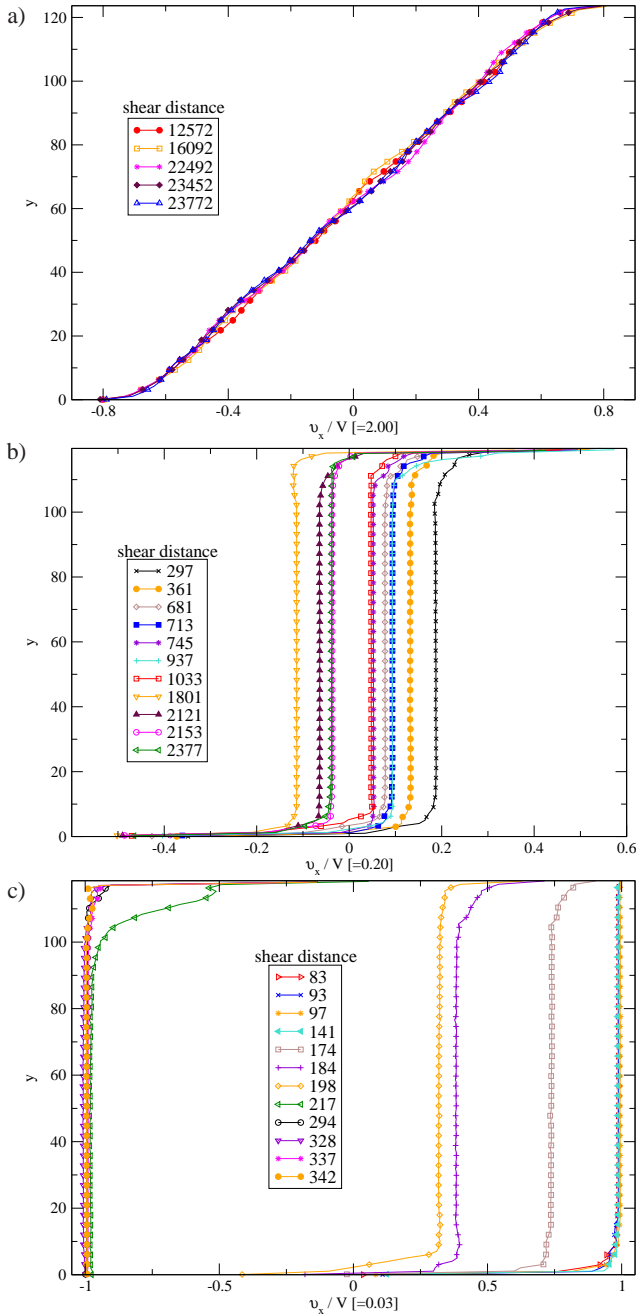

  \centering
  \raisebox{39ex}{a)}\includegraphics[width=0.94\columnwidth, clip]{./Figure-06-a.eps}\\
  \raisebox{39ex}{b)}\includegraphics[width=0.94\columnwidth, clip]{./Figure-06-b.eps}\\
  \raisebox{39ex}{c)}\includegraphics[width=0.94\columnwidth, clip]{./Figure-06-c.eps}

  \caption{(color online) Velocity profiles at different shear
    distances for three different values of $V$, as indicated, in
    sample with height $L_y{=}120$ (System $6$ in
    Tab.~\ref{tab:LxLy}).}
    \label{fig:velocity-profile}
\end{figure}

Fig.~\ref{fig:velocity-profile} displays the time evolution of the
velocity profiles of a system of initial height $L_y{=}120$ sheared
with different velocities (a)~$V{=}2.0$, (b)~$V{=}0.2$ and
(c)~$V{=}0.03$. Those three cases are characteristics of three different
regimes observed in different intervals as velocity $V$ decreases. At
large velocity $V$, as for $V{=}2.0$ (panel~(a) in
Fig.~\ref{fig:velocity-profile}), the velocity profile adopts (after a
transient) a symmetric, almost linear shape with only small
fluctuations in time. The shear rate is somewhat larger near the walls
than in the bulk, but the latter region is homogeneously sheared. This
is {\em the fast or homogeneous shear regime} (regime A in the
sequel). For intermediate velocities, such as $V{=}0.2$
(panel~(b) in Fig.~\ref{fig:velocity-profile}) the
shear rate is strongly localized at the walls, while, ten grain
diameters away from the walls, the material is hardly sheared at
all. While the profile shape is essentially stable, its position on
the velocity axis fluctuates notably: the bulk material behaves like a
solid block, but its velocity exhibits large fluctuations. To this
situation we shall refer as {\em the intermediate or two-shear band
  regime} (regime B). Finally, for a low enough shear velocity, as for
$V{=}0.03$ (panel~(c) in
Fig.~\ref{fig:velocity-profile}) the profiles fluctuate very strongly
and the top/bottom symmetry is broken. The shear strain strongly
localizes at one wall, while the rest of the system, the bulk region
and the opposite wall, moves like one single solid
object~\cite{Shojaaee_07, Shojaaee_etal_07, Shojaaee_etal_09}. This is
{\em the slow shear or one-shear band regime} (regime C). Localization
occasionally switches to the other wall, with a transition time that
strongly depends on the system size and on the shear velocity (as we
shall discuss in Sec.~\ref{subsec:Transient Time}).

In regime A the sheared layer behaves similarly to the observations
reported by da Cruz \emph{et al.}~\cite{daCruz_etal_05}, in a
numerical study of steady uniform shear flow of a granular material
between rough walls. However, with rough walls the homogeneous shear
regime persists down to very low velocities, in spite of increasing
fluctuations. The smooth walls of our system, allowing for slip and
rotation at the walls, are responsible for the more complex behavior
\cite{Shojaaee_07, Shojaaee_etal_07, Shojaaee_etal_09}.

In order to be able to observe the three regimes, sample height $L_y$
should be large enough. In smaller systems ($L_y\,{\lesssim}\,80$) the
effects of the boundary layers on the central region are strong enough
to preclude the observation of a clearly developed intermediate
regime. Sheared granular layers of smaller thickness most often
exhibit a direct transition from regime A to regime C on decreasing
velocity $V$.

Our system size analysis shows a discontinuous transition from regime
B to regime C, at $\vcone{\simeq}0.10$ and a continuous transition
between regimes A and B completed at $\vctwo{\simeq}0.50$
\cite{Shojaaee_07, Shojaaee_etal_07, Shojaaee_etal_09}. $\vcone$ and
$\vctwo$ are system size independent.

\begin{figure}[t!]
  \centering
  \includegraphics[width=\columnwidth, clip]{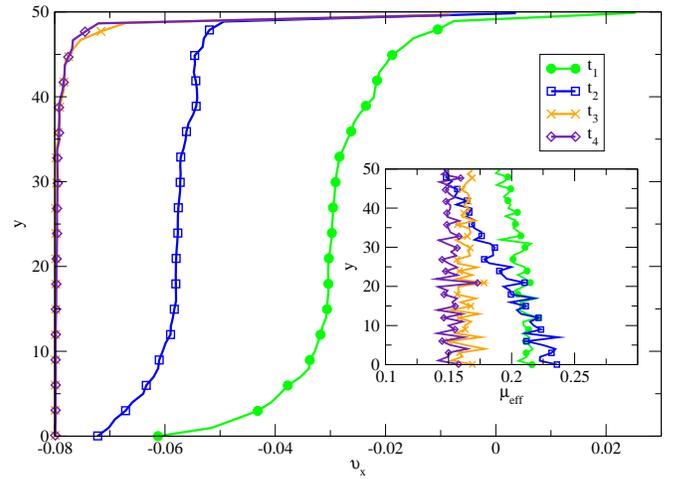}
  \caption{(color online) Profiles of velocity and effective friction
    coefficient (inset) in steady state and in the transient states
    for $V{=}0.08$ in a system with $L_x{=}50$ and $L_y{=}50$.}
  \label{fig:SS-V0.08}
\end{figure}

Upon reducing the shear velocity in the intermediate shear regime
towards $\vcone$ larger and larger fluctuations in the velocity fields
are observed, involving increasingly long correlation times. Slightly
above $\vcone$ the approach to a steady state becomes problematic,
even after the largest simulated shear strain (or wall displacement)
intervals. Then below $\vcone$ the width of the distribution of the
bulk region velocities reaches its maximum value, $2V$, and the
velocity profile stays for longer and longer time intervals in the
localized state with one shear band at a wall (regime C). Such
localized profiles can be regarded as quasi-steady states -- as
switches from one wall to the opposite one, ever rarer at lower
velocities, sometimes occur. The lifetime of these one-shear band
asymmetric steady shear profiles also increases with system height
$L_y$, similar to ergodic time in magnetic systems
\cite{Shojaaee_07,Shojaaee_etal_07}. These quasi-steady states also
exhibit uniform stress profiles, contrary to the nonuniform ones in
the transient states, as the localization pattern is switching to the
other side (Fig.~\ref{fig:SS-V0.08}).

\begin{figure}[t]
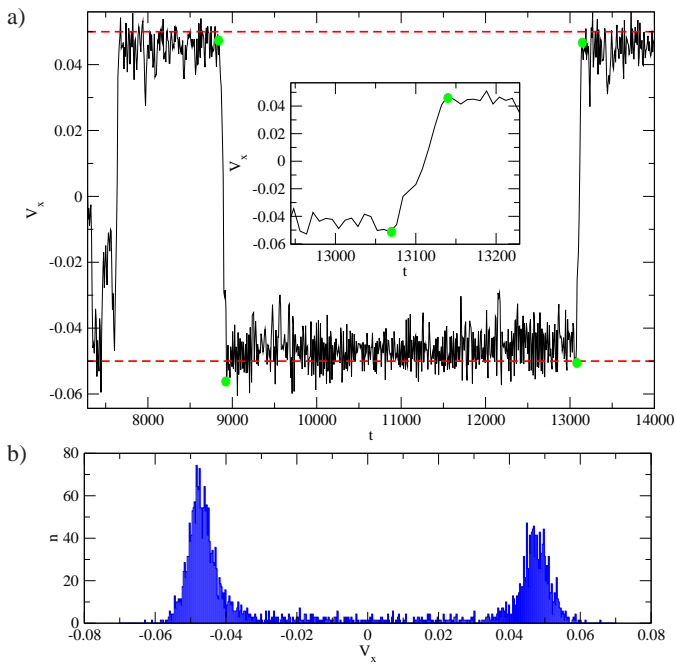

  \centering
  \raisebox{39ex}{a)}\includegraphics[width=\columnwidth, clip]{./Figure-08-a.eps}
  \raisebox{19ex}{b)}\mbox{\hspace{2ex}\includegraphics[width=0.95\columnwidth, clip]{./Figure-08-b.eps}}
  \caption{(color online) (a) Center of mass velocity fluctuations in
    steady state for $V{=}0.05$ in a system with $L_x{=}20$ and
    $L_y{=}20$. The dashed red lines represent the velocity of the top
    and bottom walls. The transition time (magnified in the inset) is
    measured at both ends of direct transitions from one wall to the
    other, between the round dots. (b) Histogram of center of mass
    velocities.}
  \label{fig:V.x-t-0.05-Ly20}
\end{figure}

Fig.~\ref{fig:V.x-t-0.05-Ly20}~(a) is a plot of center of mass velocity in
the flow direction versus time in regime C. Most of the time, it is
slightly fluctuating about the value of either one of the velocities
of the walls, $\pm V$, as also illustrated in the histogram plot,
Fig.~\ref{fig:V.x-t-0.05-Ly20}~(b), for which values were accumulated
over a long time and over different simulated systems. Transition
times as the shear band switches directly from one wall to the other
are measured as indicated. Those times are recorded to be discussed in
Sec.~\ref{subsub:Transient from one to the other Wall in Regime c}.

\subsection{Slip velocity}
\label{subsection:Slip Velocity}
\index{Slip Velocity|(}

The slip at smooth walls is a characteristic feature of the boundary
region behavior.

To evaluate the slip velocity at the walls one needs to calculate the
average of the surface velocity of particles in contact with the walls
at their contact point. The slip velocity in this work is defined as
the absolute value of the difference between the wall velocity and the
average particle surface velocity at the corresponding wall,
$\upsilon_\text{0}^\text{slip}$ at the bottom, respectively
$\upsilon_{\text{L}_y}^\text{slip}$ at the top wall. To this end all
particles in contact with the walls over the whole simulation time in
steady state should be considered, and contribute
\begin{gather}\label{v.x}
  \upsilon_\text{0}^\text{slip}=V + \langle \upsilon_{ix}
  + \omega_i r_i \rangle_{i ,
    t}, \\
  \upsilon_{\text{L}_y}^\text{slip}=V - \langle \upsilon_{ix}
  - \omega_i r_i \rangle_{i ,
    t},
\end{gather}
where $\upsilon_{ix}$ is the $x$ component of the center of mass
velocity of particle $i$ of radius $r_i$ with angular velocity
$\omega_i$.

\begin{figure}[t]
  \centering
  \includegraphics[width=1\columnwidth, clip]{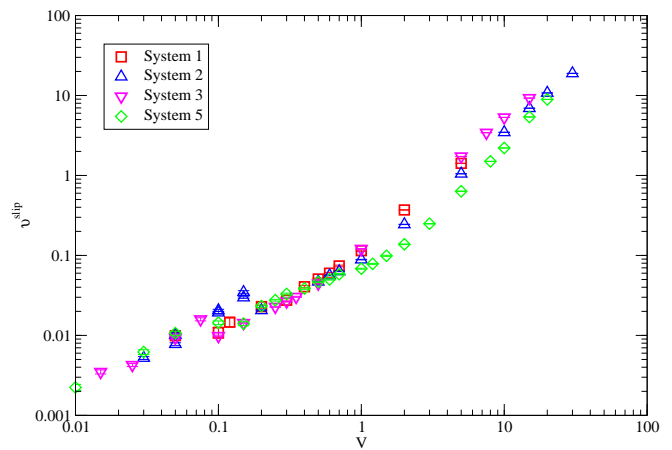}
  \caption{(color online) Slip velocity $\upsilon^\text{slip}$
    (averaged over $\upsilon_\text{0}^\text{slip}$ and
    $\upsilon_{\text{L}_y}^\text{slip}$) measured as a function of shear
    velocity (systems specified in Tab.~\ref{tab:LxLy}).}
  \label{fig:vslip-V}
\end{figure}

Our observations show that the slip velocity in a certain shear
velocity interval $0.2\,{\lesssim}\,V\,{\lesssim}\,1.0$ does not depend on the
system size (Fig.~\ref{fig:vslip-V}). For larger shear velocities,
though the general tendency is the same, slight deviations are
observable. The apparent change of slope of the graph near $V{=}1$
could be associated with larger strain rates and inertial numbers in
boundary layers, gradually approaching a collisional regime (see
Sec.~\ref {subsec:Coordination number}, about the coordination
number).

\section{Constitutive laws}
\label{sec-constitutive laws}

Constitutive laws were previously studied, in similar model materials,
in homogeneous shear flow~\cite{daCruz_etal_05, Hatano_07,
  Peyneau-Roux_08}. Their sensitivity to material parameters
(restitution coefficients, friction coefficients and, possibly, finite
contact stiffness) is reported, e.g. in~\cite{RoCh11}. In our system
we separate the boundary regions near both walls, from the central one
(or bulk region). Unless otherwise specified, the boundary regions
have thickness $h=10$. Near the walls, the internal state of the
granular material is different, and we seek separate constitutive laws
for the boundary layers and for the bulk material. While the bulk
material is expected to abide by constitutive laws that apply locally,
and should be the same as the ones identified in other geometries or
with other boundary conditions~\cite{daCruz_etal_05,Koval_etal_09},
the boundary constitutive law is expected to relate stresses to the
global velocity variation across the layer adjacent to the wall. In a
continuum description suitable for large scale problems, this will
reduce to relating stresses to tangential velocity discontinuity.

\subsection{Constitutive laws in the bulk region}
\label{subsec-constitutive laws in bulk}

\subsubsection{Friction law}
\label{subsub:Friction Law bulk}

\begin{figure}[t]
  \centering
  \includegraphics[width=1\columnwidth, clip]{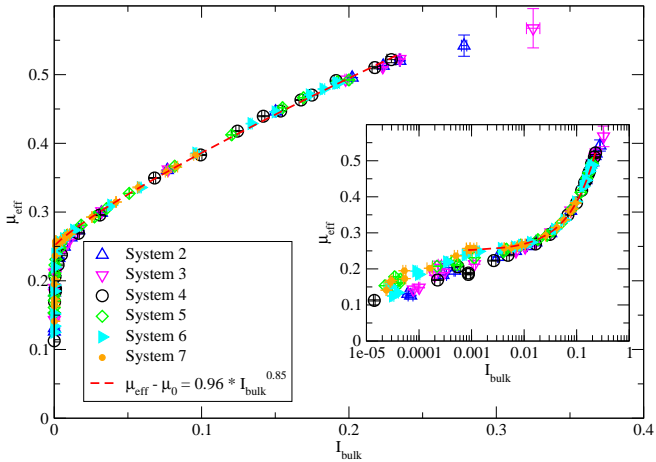}
  \caption{(color online) \muf\ as a function of inertial number in
    the bulk region for different system sizes (see
    Tab.~\ref{tab:LxLy}). The fit function is calculated according to
    Eq.~\ref{mu_I-non-linear} for $\mu_0{=}0.25$. The error bars are
    much smaller than the symbols. The inset is a semilogarithmic plot
    of the same data.}
  \label{fig:mu-I-bulk}
\end{figure}

The steady state values of the inertial number ($I_\text{bulk}$) and
that of the effective friction coefficient \muf\ are measured, as
averages over time and over coordinate $y$ within the interval
$h<y<L_y-h$. \muf\ is plotted as a function of $I_\text{bulk}$ for all
different system sizes in Fig.~\ref{fig:mu-I-bulk}, showing data
collapse for different sample sizes.

\begin{figure}[t]
  \centering
  \includegraphics[width=1\columnwidth, clip]{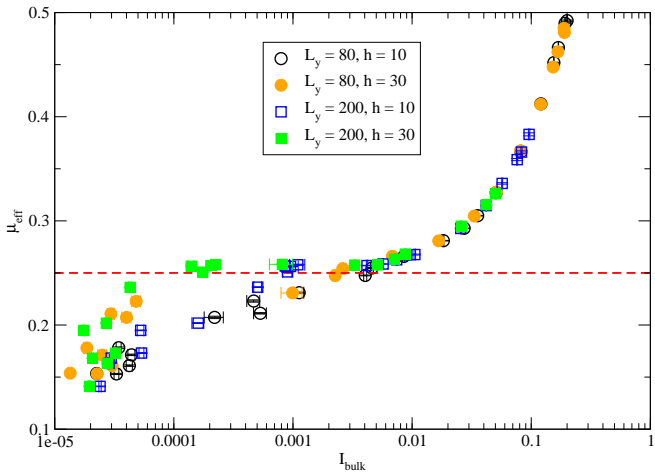}
  \caption{(color online) Influence of $h$ on \muf\ as a function of
    inertial number in the bulk region (data from systems $5$ and $7$
    in Tab.~\ref{tab:LxLy}). The dashed red line represents the
   critical friction coefficient $\mu_0{=}0.25$.}
  \label{fig:mu-I-bulk-h}
\end{figure}

The apparent influence of the choice of $h$ on the measured effective
friction coefficient and inertial number in the bulk region is
presented in Fig.~\ref{fig:mu-I-bulk-h} for two different system sizes
and for two different $h$ values.

We observe that some data points with finite values of $I_\text{bulk}$
($I_\text{bulk} > 10^{-4}$) are shifted to much smaller values of
$I_\text{bulk}$ upon increasing $h$: compare the open and full symbols
in Fig.~\ref{fig:mu-I-bulk-h}. This effect is apparent in regimes B
and C. It is due to the creep phenomenon (as was also observed in the
annular shear cell in \cite{Koval_etal_09}), which causes some amount
of shearing at the edges of the bulk region, adjacent to the boundary
layer, although the effective friction coefficient is below the
critical value. Although the local shear stress is too small for the
material to be continuously sheared, the ambient noise level, due to
the proximity of the sheared boundary layer, entails slow
rearrangements that produce macroscopic shear \cite{Unger_10}.  Upon
increasing $h$ the central bulk region excludes the outer zone that is
affected by this creep effect. The critical friction coefficient, from
Fig.~\ref{fig:mu-I-bulk-h}, is $\mu_0{=}0.25$ (below which the data
points are sensitive to the value of $h$), which is consistent with
the results of the literature~\cite{daCruz_etal_05,Koval_etal_09}.

Fitting $\Muf - \mu_0$ with a power law function, as in~\cite{Hatano_07,Peyneau-Roux_08}
\begin{gather}\label{mu_I-non-linear}
\Muf - \mu_0 = A \cdot I_\text{bulk}^{B},
\end{gather}
the following coefficient values yield good results (see Fig.~\ref{fig:mu-I-bulk}):
\begin{gather*}
  \mu_0=0.24 \pm 0.01,\\
  A=0.92 \pm 0.05,\\
  B=0.80 \pm 0.05.
\end{gather*}

\subsubsection{Dilatancy law}
\label{subsub:Dilatancy Law bulk}
\begin{figure}[t!]
  \centering
  \includegraphics[width=1\columnwidth, clip]{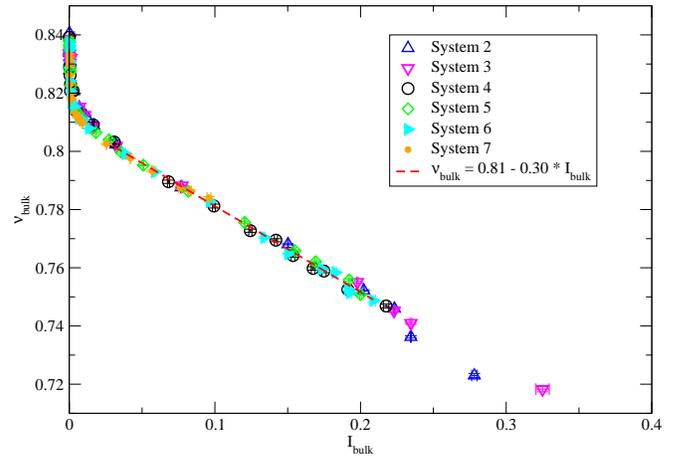}
  \caption{(color online) $\nu$ as a function of inertial number in
    the bulk region.  The error bars plotted are much smaller than the
    symbols (systems specified in Tab.~\ref{tab:LxLy}).}
  \label{fig:nu-I-bulk}
\end{figure}

We now focus on the variation of solid fraction $\nu$ as a function of
inertial number within the bulk region. $\nu$ is averaged over time,
once a steady state is achieved, within the central region,
$h<y<L_y-h$. Function $\nu_\text{bulk}(I_\text{bulk})$ is plotted in
Fig.~\ref{fig:nu-I-bulk} for different system sizes, leading once
again to a good data collapse. A linear fit for all data sets in the
interval $0.03 < I_\text{bulk} < 0.20$ gives:
\begin{gather}\label{nu_I}
\nu_\text{bulk} = 0.81 - 0.30 \cdot I_\text{bulk},  
\end{gather}
which is consistent with the linear fit in
\cite{daCruz_etal_05,Koval_etal_09}.

\subsection{Constitutive laws in the boundary layer}
\label{subsec-constitutive laws in boundary}
In order to characterize the state of the boundary layer of width $h$
adjacent to the wall (recall $h=10$ by default), we use a local
inertial number $I_\text{boundary}$, defined as follows:
\begin{gather}\label{I_boundary}
  I_\text{boundary}^\text{top/bottom}=\sqrt{\dfrac{m}{\sigma_{yy}}}\times \left <
    \dfrac{\Delta \upsilon^\text{top/bottom}}{h} \right >_{t},
\end{gather}
with 
\begin{gather}
 \Delta \upsilon^\text{top}=V - \upsilon_x(L_y-h), \label{deltaV}\\
 \Delta \upsilon^\text{bottom}=\upsilon_x(h) + V. \notag
\end{gather}

\subsubsection{Friction law}
\label{subsub:Friction Law boundary}
\begin{figure}[t]
  \centering
  \includegraphics[width=1\columnwidth, clip]{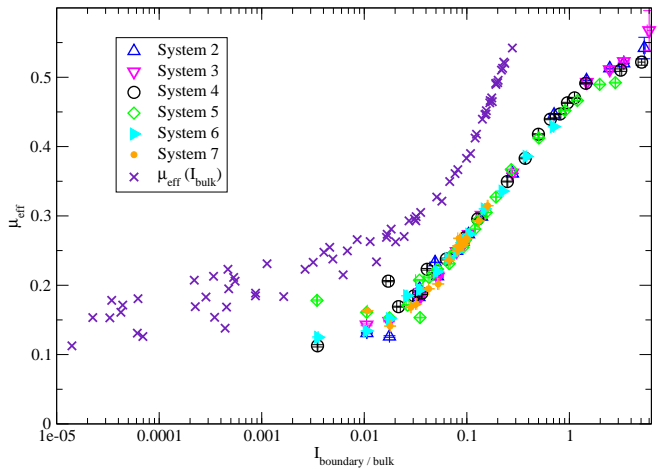}
  \caption{(color online) \muf\ as a function of inertial number in
    the boundary layer. The error bars plotted are much smaller than
    the symbols. As $I_\text{boundary}{>}I_\text{bulk}$ (shear
    localization at smooth walls) $\Muf(I_\text{boundary})$ lies
    always beneath $\Muf(I_\text{bulk})$. The
      presented $\Muf(I_\text{bulk})$ curve is based on
the same data set as Fig.~\ref{fig:mu-I-bulk}.} 
  \label{fig:mu-I-boundary}
\end{figure}

\index{Friction Law boundary|(}
Fig.~\ref{fig:mu-I-boundary} is a plot of 
\muf\ as a function of the inertial number
$I_\text{boundary}$ in the boundary layer for all different system
sizes. 

In steady state the value of \muf\ in the boundary layer has to be
equal to the averaged one in the bulk. The observed shear increase (in
regime A) or localization (in regimes B and C) near the smooth walls
entails larger values of inertial numbers in the boundary region. An
equal value of \muf\ in the bulk and in the boundary zone then
requires that the graph of function $\Muf(I_\text{boundary})$ is below
its bulk counterpart in the inertial number interval measured.

\begin{figure}[ht]
  \centering
  \includegraphics[width=\columnwidth,clip]{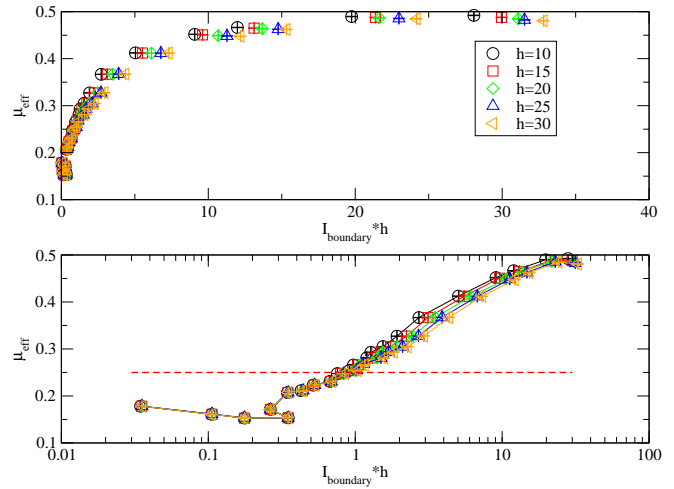}
  \caption{(color online) \muf\ versus $h\times I_\text{boundary}$ on
    linear (top panel) and semi-logarithmic (bottom panel) plots. The
    dashed horizontal line indicates the critical state value
    $\Muf=\mu_0{=}0.25$.}
  \label{fig:mu-times-h-boundary}
\end{figure}

In Sec.~\ref{subsec-constitutive laws in bulk} we have seen that the
friction law can be identified in the bulk independently of $h$ (see
Fig.~\ref{fig:mu-I-bulk-h}), as an intrinsic constitutive law.
According to the definition of $I_\text{boundary}$ in
Eqs.~(\ref{I_boundary}) and (\ref{deltaV}) any constitutive relation
involving $I_\text{boundary}$ should trivially depend on $h$. In shear
regimes B and C, there is no shearing in the bulk region, and
consequently $\Delta \upsilon$ in the numerator of
Eq.~(\ref{I_boundary}) does not change with $h$. On multiplying the
measured $I_\text{boundary}$ with the corresponding value of $h$, we
thus expect the data points belonging to shear regimes B and C to
coincide (Fig.~\ref{fig:mu-times-h-boundary}). In regime A, in
contrast, the existence of shear in the bulk region leads to an
apparent $h$ dependence of the measured $\Delta
\upsilon$. Accordingly, after multiplying $I_\text{boundary}$ with
$h$, the curves do not merge. The critical effective friction
coefficient at which the deviation of the curves begins corresponds to
$\mu_0{=}0.25$ (the dashed horizontal line in
Fig.~\ref{fig:mu-times-h-boundary}), in agreement with the results in
Sec.~\ref{subsub:Friction Law bulk}. This makes it more difficult to
identify a constitutive law for the boundary layer, when the bulk
region is sheared in regime A.

\begin{figure}[t]
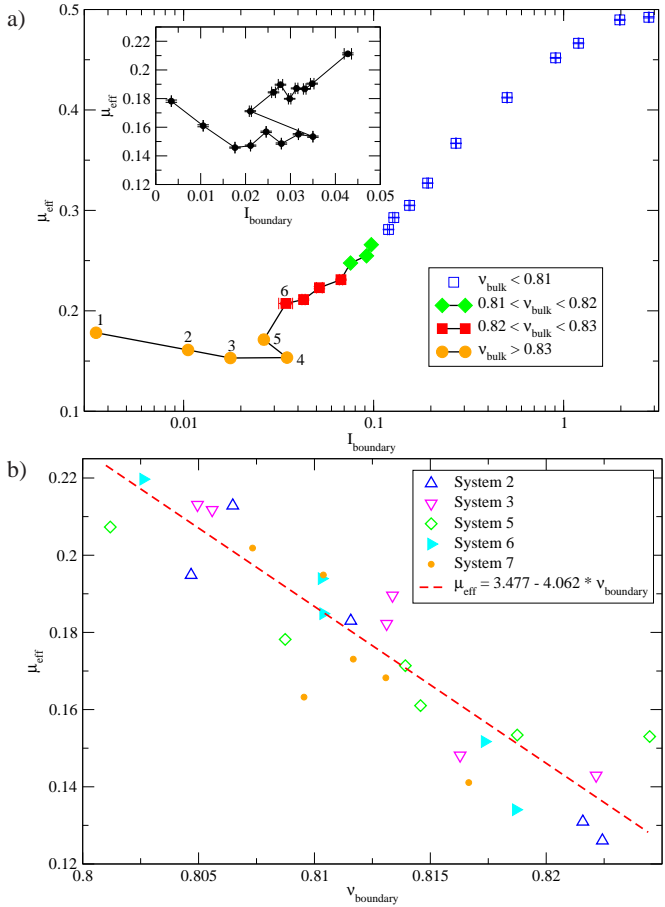

  \centering
  \raisebox{40ex}{a)}\mbox{\hspace{1ex}\includegraphics[width=0.96\columnwidth, clip]{./Figure-15-a.eps}}\\
  \raisebox{40ex}{b)}\includegraphics[width=0.98\columnwidth, clip]{./Figure-15-b.eps}\\
  \caption{(color online) (a) $\mu_\text{eff}$ as a function of
    $I_\text{boundary}$ (data from system $5$).The full symbols belong
    to the states with bulk densities larger than the critical value
    $\nu_c{=}0.81$ (see Eq.~(\ref{nu_I})). The full diamonds have a
    density between $0.81$ and $0.82$, full squares have a density
    between $0.82$ and $0.83$ and full circles have a density larger
    than $0.83$. The inset represents the zig-zag with some more data
    points, which are absent in the master graph for the sake of
    clarity. (b) $\mu_\text{eff}$ as a function of
    $\nu_\text{boundary}$. The error bars are smaller than the
    symbols.}
  \label{fig:mu-eff-nuC}
\end{figure}

The behavior of \muf\ shown in Fig.~\ref{fig:mu-times-h-boundary} is
apparently anomalous in two respects: $(i)$ the $\Delta \upsilon$
dependence of \muf\ does not seem to follow a single curve (suggesting
\muf\ depends on other state parameters than the velocity variation
across the boundary zone); $(ii)$ \muf\ is a decreasing function of
$I_\text{boundary}$ for the first data points, as $h\times
I_\text{boundary} < 0.2$. In Fig.~\ref{fig:mu-eff-nuC}~(a), we take a
closer look at the low $I_\text{boundary}$ data points, which bear
number labels 1 to 6 in the order of increasing shear velocity
$V$. The transition from regime C (one shear band) to regime B (two
shear bands) occurs between points 4 and 5, whence a decrease in
$I_\text{boundary}$, as the velocity change across the sheared
boundary layers changes from $2V$ to merely $V$. In an attempt to
identify one possible other variable influencing boundary layer
friction, the symbols on Fig.~\ref{fig:mu-eff-nuC}~(a) also encode the
value of the bulk density. We note then that points 4 and 6, which
have different friction levels, although approximately the same
$I_\text{boundary}$, correspond to different bulk densities. The
constitutive laws in the boundary layer might thus depend on parameter
$\nu_\text{bulk}$ in addition to $I_\text{boundary}$.

As to issue $(ii)$, the decrease of \muf\ before the zig-zag pattern
on the curve of Fig.~\ref{fig:mu-eff-nuC}~(a) (data points $1$ to $3$)
is associated to an increase in the boundary layer density with
$I_\text{boundary}$. This is not the case in all of the systems and
these features strongly depend on the preparation and the initial
packing density (compaction in the absence of friction). Independent
of whether \muf\ in regime C increases or not as $I_\text{boundary}$
increases, \muf\ always displays a decreasing tendency as
$\nu_\text{boundary}$ increases, just like \muf\ and $\nu$ vary in
opposite directions in bulk systems under controlled normal stress, as
shown in Ref.~\cite{daCruz_etal_05}, or as expressed by
Eqs.~\eqref{mu_I-non-linear} and \eqref{nu_I} (panel~(b) in
Fig.~\ref{fig:mu-eff-nuC}). The lack of a perfect collapse of the data
points around the decreasing linear fit of
Fig.~\ref{fig:mu-eff-nuC}~(b) shows however that the state of the
boundary layer in slowly sheared systems does not depend on a single
local variable, but is influenced by the state of the neighboring bulk
material, as remarked above.

\subsubsection{Dilatancy law}
\label{subsub:Dilatancy Law boundary}
\index{Dilatancy Law boundary|(} 

\begin{figure}[t!]
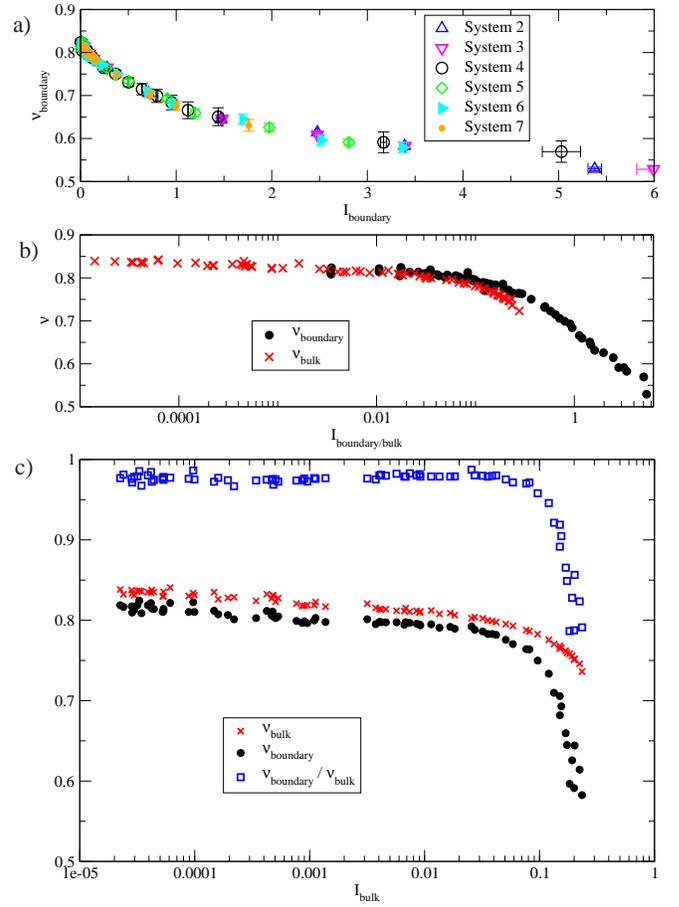

  \centering
  \raisebox{18ex}{a)}\mbox{\hspace{0ex}\includegraphics[width=0.97\columnwidth, clip]{./Figure-16-a.eps}}\\
  \raisebox{18ex}{b)}\mbox{\hspace{0.0ex}\includegraphics[width=0.948\columnwidth, clip]{./Figure-16-b.eps}}\\
  \raisebox{40ex}{c)}\mbox{\hspace{1.8ex}\includegraphics[width=0.933\columnwidth, clip]{./Figure-16-c.eps}}\\
  \caption{(color online) (a) $\nu$ as a function of inertial number in
    the boundary layers (systems specified in Tab.~\ref{tab:LxLy}). (b)
    $\nu_\text{boundary}(I_\text{boundary})$ compared to
    $\nu_\text{bulk}(I_\text{bulk})$. (c)~The ratio between
    $\nu_\text{boundary}$ and $\nu_\text{bulk}$ as a function of
    $I_\text{bulk}$.}
  \label{fig:nu-I-boundary}
\end{figure}

After averaging the profiles of solid fraction and inertial number
over the whole simulation time in steady state in the boundary region,
$\nu_\text{boundary}(I_\text{boundary})$ graphs are then plotted in
Fig.~\ref{fig:nu-I-boundary}~(a) for different system sizes. In
Fig.~\ref{fig:nu-I-boundary}~(b), $\nu_\text{bulk}(I_\text{bulk})$ and
$\nu_\text{boundary}(I_\text{boundary})$ are compared for all data
sets. $\nu_\text{boundary}$ and $\nu_\text{bulk}$ drop proportionally
with increasing $I_\text{bulk}$ (shear velocity) until
$I_\text{bulk}{\simeq}0.08$ (in shear regime A). Afterwards, the drop
in $\nu_\text{boundary}$ is much steeper
(Fig.~\ref{fig:nu-I-boundary}~(c)).

\subsection{Coordination number}
\label{subsec:Coordination number}
\index{Coordination number} 

\begin{figure}[t]
  \centering
    \includegraphics[width=\columnwidth , clip]{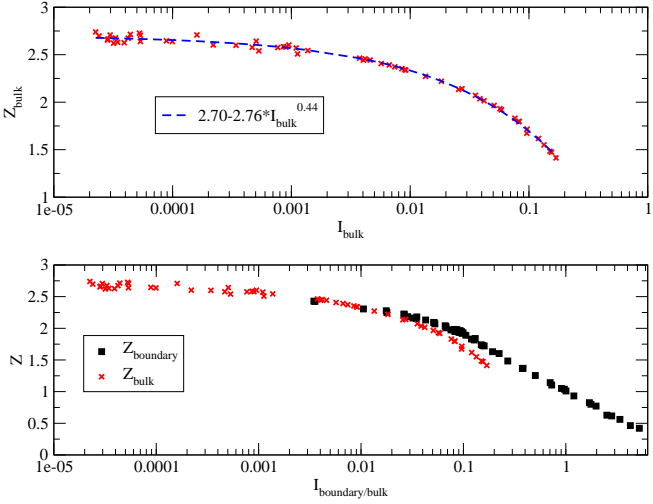}
    \caption{(color online) Coordination number as a function of
      inertial number in the bulk and boundary regions. The error bars
      are much smaller than the symbols.}
  \label{fig:co-I}
\end{figure}

The coordination number (average number of contacts per grain) is a
quantitative measure of the status of the contact
network. Fig.~\ref{fig:co-I} shows the measured coordination number in
the bulk and in the boundary layers as a function of inertial numbers
in these two regions. The data are collected from different systems in
Tab.~\ref{tab:LxLy}.

The bulk coordination number, $Z_\text{bulk}$ is fitted with the power
law function $Z_\text{bulk}=2.70-2.76 \cdot I_\text{bulk}^{0.44}$. The
boundary region coordination number, $Z_\text{boundary}$, follows a
slightly different dependency on $I_\text{boundary}$, which becomes
noticeable for $I_\text{boundary}\,{\gtrsim}\,0.1$, which corresponds to
$V{\simeq}1$ (see Sec.\ref{subsection:Slip Velocity}). It drops to
smaller values, as $I_\text{boundary}$ reaches larger values, above
1. The decrease of coordination number as a function of inertial
number is compatible with the observations of \cite{daCruz_etal_05},
in which some effect of restitution coefficient on $Z$ was however
reported. The finite softness of the particles is also known to affect
coordination numbers~\cite{Silbert_etal_01} much more than the
rheological laws. It is only for configurations extremely close to
equilibrium that coordination numbers are observed to exceed the
minimum value 3 for stable packings of frictionless disks (excluding
rattlers), and even in quite slow flows this condition is not
fulfilled.

\section{Applications}
\label{sec:Applications} 
We now exploit the constitutive relations and other observations
reported in the previous sections to try and deduce some features of
the global behavior of granular samples sheared between smooth walls.

\subsection{Transient time}
\label{subsec:Transient Time}

\subsubsection{Transient to steady state in regime A}
\label{subsub:Transient to Steady State in Regime a}

The bulk friction law of Sec.~\ref{subsub:Friction Law bulk} can be
used to evaluate the time for a system to reach a uniform shear rate
in regime A, if we assume constant and uniform solid fraction $\nu$
and normal stress $\sigma_{yy}$, and velocities parallel to the walls
at all times. We write down the following momentum balance equation:
\begin{gather}  
  \dfrac{\partial (\rho \nu \upsilon_x)}{\partial t} = \dfrac{\partial
    \sigma_{xy}}{\partial y},
\end{gather}
looking for the steady solution: $\upsilon_x=\dot{\gamma} y$. Assuming
constant $\rho$, $\nu$ and $\sigma_{yy}$ we can write:
\begin{gather}
  \rho \nu \dfrac{\partial \upsilon_x}{\partial t} =
  \dfrac{\partial}{\partial y}\left[\Muf(\dot{\gamma})\right]
  \sigma_{yy},
\end{gather}
which leads by derivation to:
\begin{gather}\label{eq1}
  \rho \nu  \dfrac{\partial
  \dot{\gamma}}{\partial t} = \dfrac{\partial ^2}{\partial
  y^2}\left[\Muf(\dot{\gamma}) - \mu_0\right] \sigma_{yy}.
\end{gather}
Separating the shear rate field into a uniform part $\dot{\gamma}_0$
and a $y$-dependent increment $\Delta \dot{\gamma}$, and assuming as
an approximation just a linear dependency of \muf\ on $\dot{\gamma}$,
we can rewrite Eq.~(\ref{eq1}) as follows:
\begin{gather}
  \rho \nu \dfrac{\partial \Delta \dot{\gamma}}{\partial t} = \sigma_{yy}
  \dfrac{\partial \Muf}{\partial \dot{\gamma}}\dfrac{\partial
    ^2}{\partial y^2} \Delta \dot{\gamma},
\end{gather}
which is a diffusion equation with diffusion coefficient
\begin{gather}
D=\frac{\partial \Muf}{\partial \dot{\gamma}}\frac{\sigma_{yy}}{\rho \nu}. \label{eq:defD}
\end{gather}
The characteristic time to establish the steady state profile (uniform
$\dot{\gamma}$ over the whole sample height $L_y$) is then:
\begin{gather}
  T_\text{SS} = \dfrac{L_y^2 }{D}. \label{eq:tss}
\end{gather}
A linear fit of function $\Muf(I_\text{bulk})$ (see
Fig.~\ref{fig:mu-I-bulk}) in interval ($0.03 < I_\text{bulk} < 0.20$)
is:
\begin{gather}\label{mu_I-linear}
\Muf = 0.27 + 1.16 \cdot I_\text{bulk}.  
\end{gather}
According to Eqs.~\eqref{inertial_number}, \eqref{eq:defD},
\eqref{eq:tss} and (\ref{mu_I-linear}) this leads to:
\begin{gather}\label{eq-T-SS}
  T_\text{SS} \simeq 1.56 L_y^2.
\end{gather}
The estimated values $T_\text{SS}$ for different system sizes is
listed in Tab.~\ref{tab:LxLy}. As $T_\text{SS}$ grows like $L_y^2$,
very long simulation runs become necessary to achieve steady states in
tall (large $L_y$) samples, and some unstable, but rather persistent,
distributions of shear rate can be
observed~\cite{Aharonov_Sparks_02,Peyneau-Roux_08}. Our data for
$L_y{=}120$ and $L_y{=}200$ may still pertain to slowly evolving
profiles, even though the constitutive law can be measured in
approximately homogeneous regions of the sheared layer over time
intervals in which profile changes are negligible.

\subsubsection{Transition from one wall to the other in regime C}
\label{subsub:Transient from one to the other Wall in Regime c}

\begin{figure}[t]
  \centering
    \includegraphics[width=\columnwidth , clip]{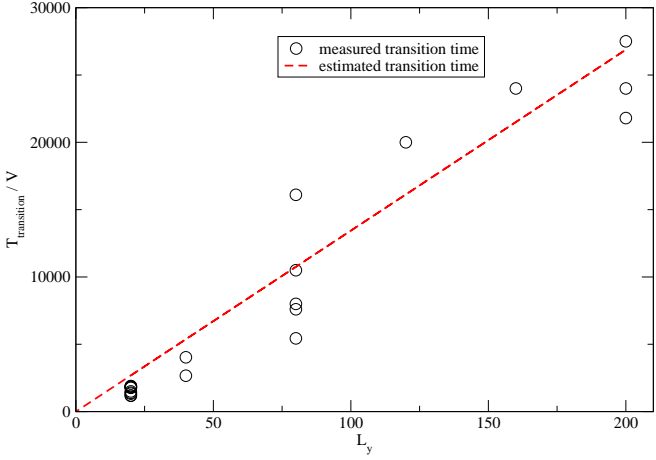}
    \caption{(color online) Transition time divided by the shear
      velocity as a function of system height. Empty circles 
correspond to measured times, while the (red) dashed line plots estimated ones, using~\eqref{eq:ttcal}.}
  \label{fig:T-Transient-Ly}
\end{figure}

As stated in Sec.~\ref{subsection:Shear Regimes} in regime C the
asymmetric velocity profiles can be regarded as steady states and the
switching stages in which the shear band changes sides are transient
states in which the shear stress is not uniform through the granular
layer. We now try to estimate the characteristic time for such
transitions. This estimation does not rely on a specific model for the
triggering mechanism of the transition. It is based on the simple idea
that the transition takes place when the solid block is accelerated
due to a shear stress difference between the top and the bottom
boundary zones. Taking the whole bulk region as a block of mass $M$
moving with the velocity of the top wall $V$, a transition to velocity
$-V$ with acceleration $A$ will take:
\begin{gather}
  T_\text{transition}=\dfrac{2V}{A},
\end{gather}
in which the acceleration $A$ is equal to:
\begin{gather}
  A=\dfrac{(\sigma_{xy}^{top}-\sigma_{xy}^{bottom})L_x}{M}.
\end{gather}
Substituting $M{=}\rho \nu L_x L_y$ and
$\sigma_{xy}^{top}-\sigma_{xy}^{bottom}{=}\Delta \mu \sigma_{yy}$ with
$\Delta \mu {=} \mu^{top} - \mu^{bottom}$ one gets:
\begin{gather}
  T_\text{transition}=\dfrac{2 \rho \nu V L_y}{\Delta \mu \sigma_{yy}}.\label{eq:ttcal}
\end{gather}
Accordingly, the transition time increases proportionally to the shear
velocity and to system height $L_y$. Using $\nu \simeq 0.84$,
$\sigma_{yy} = 0.25$ and taking $\Delta \mu \simeq 0.05$ as a
plausible value in shear regime C (see Figs.~ \ref{fig:SS-V0.08} and
\ref{fig:mu-times-h-boundary}) we calculate
$\frac{T_\text{transition}}{V}$ as a function of system height
$L_y$. In Fig.~\ref{fig:T-Transient-Ly} these estimated times are
compared to transition times that are measured as explained in the
caption of Fig.~\ref{fig:V.x-t-0.05-Ly20}.

Admittedly, one does not observe only direct, sharp transitions in
which localization changes from one wall to the opposite one. Some
transient states are more uncertain and fluctuating, and the system
occasionally returns to a localized state on the same wall after some
velocity gradient has temporarily propagated within the central
region. The data points of Fig.~\ref{fig:T-Transient-Ly} correspond to
the well-defined transitions, which become less frequent with
increasing system height. Thus a unique data point was recorded for
systems with $L_y=120$ and $L_y=160$. The comparison between estimated
and measured transition times is encouraging, although the value of
$\Delta \mu$ in \eqref{eq:ttcal} is of course merely indicative (it is
likely to vary during the transition), and the origin of such
asymmetries between walls is not clear.

\subsection{Transition velocity $\vctwo$}
\label{subsec:Transition Velocity}

$\mu_0{=} 0.25$ from the power law fit in Eq.~(\ref{mu_I-non-linear})
corresponds to the minimal value of the bulk effective friction
coefficient, the critical value below which the granular material
cannot be continuously sheared (except for local creep effects in the
immediate vicinity of an agitated layer).

\begin{figure}[t]
  \centering
    \includegraphics[width=\columnwidth , clip]{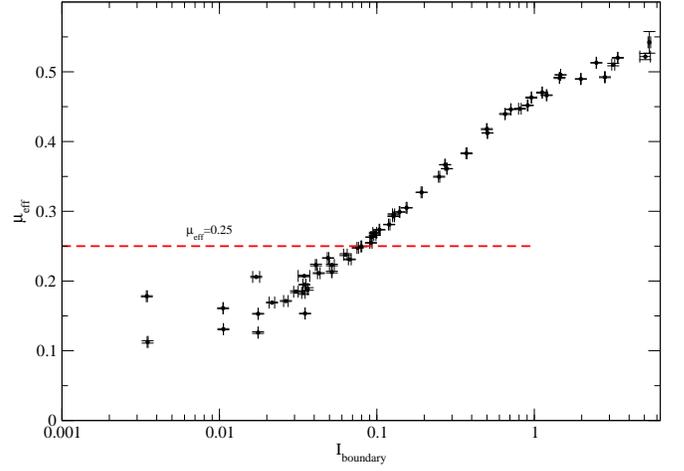}
    \caption{(color online) The critical $I_\text{boundary}$, which
      corresponds to $\mu_0{=} 0.250$ (dashed red line) and determines
      the the critical velocity $\vctwo$ for the transition from
      regime A to regime B.}
  \label{fig:mu-bulk-min-intersection}
\end{figure}

Fig.~\ref{fig:mu-bulk-min-intersection} gives the value of the
inertial number in the boundary region, such that the boundary
friction coefficient matches $\mu_0{=} 0.25$:
\begin{gather} \label{mu-min}
\mu_0 = 0.25 \Rightarrow I_\text{boundary}{=}0.086 \pm 0.005.
\end{gather}
Thus for $I_\text{boundary}\,{\lesssim}\,0.086$ we expect no shearing in
the bulk. According to Eqs.~\eqref{I_boundary} and \eqref{deltaV} this
results in $V{=}0.485 \pm 0.028$, in very good agreement with our
observations reported in Sec.~\ref{subsection:Shear Regimes}
($\vctwo\simeq 0.50$).

The explanation of the transition from regime A to regime B is simple:
the boundary layer, with a smooth, frictional wall, has a lower shear
strength (as expressed by a friction coefficient) than the bulk
material. Thus for uniform values of stresses $\sigma_{yy}$ and
$\sigma_{xy}$ in the sample, such that their ratio
$\sigma_{xy}/\sigma_{yy}$ is comprised between the static friction
coefficient of the bulk material and that of the boundary layers,
shear flow is confined to the latter.

\subsection{Transition to regime C at velocity $\vcone$}
\label{subsec:Transition Velocity 1}

Although it is not systematically observed, and is likely to depend on
the bulk density, the decreasing trend of \muf\ in the boundary layer
as a function of $\Delta \upsilon$ or of $I_{\text{boundary}}$, as
apparent in Figs.~\ref{fig:mu-times-h-boundary} and
\ref{fig:mu-eff-nuC}~(a), provides a tempting explanation to the
transition from regime B to regime C.  Assuming \muf\, for given,
constant $\sigma_{yy}$, to vary in the boundary layers as
\begin{gather}
\Muf = \mu_0 -\alpha \vert \Delta \upsilon \vert,\ \ \mbox{with} \ \alpha>0,
\end{gather}
one may straightforwardly show that the symmetric solution with
$\Delta \upsilon = \pm V$, and solid bulk velocity $\upsilon_s = 0$,
is unstable. A simple calculation similar to the one of
Sec.~\ref{subsub:Transient from one to the other Wall in Regime c}
shows that velocity $\upsilon_s$, if it differs from zero by a small
quantity $\delta \upsilon_s$ at $t=0$, will grow exponentially,
\begin{gather}
\upsilon_s(t) = \delta \upsilon_s \exp \frac{2\alpha L_x \sigma_{yy} t}{M},
\end{gather}
until it reaches $\pm V$, with the sign of the initial perturbation
$\delta \upsilon_s$. Transition velocity $\vcone$ would then be
associated to a range of velocity differences $\Delta \upsilon$ across
the boundary layer with softening behavior (i.e., decreasing function
$\Muf(I_{\text{boundary}})$).

In view of Fig.~\ref{fig:mu-eff-nuC}~(a), where the BC transition
takes place at point 4, this seems plausible, as the slope of function
$\Muf(I_{\text{boundary}})$ appears to vanish towards this point.

\section{Conclusion}
\label{sec-Conclusion-and-Discussion}

In this work we have investigated shear localization at smooth
frictional walls in a dense sheared layer of a model granular
material. The slip at the walls induces inhomogeneities in the system
leading to three different shear regimes. As the wall velocity is
reduced from large values, two transitions successively occur, in
which shear deformation localizes, first symmetrically near opposite
walls, and then at a single wall. Measuring stress tensor, inertial
number and solid fraction locally in the whole system, the
constitutive laws have been identified in the bulk (for which our
results agree with the published literature) and in the boundary
layer. Those constitutive laws, supplemented by an elementary
stability analysis, allow us to predict the occurrence of both
transitions, as well as characteristic transient times. The
consistence of the derived constitutive laws for the bulk rheology
with those in previous contributions \cite{daCruz_etal_05} using the
MD method, confirm that the rheology is the same for CD and for MD in
the limit of large contact stiffness.

Additional numerical work should be carried out in order to assess the
dependence of the boundary layer constitutive law on the state of the
adjacent bulk material with full generality. The application of
similar constitutive laws for smooth boundaries should be attempted in
a variety of flow configurations: inclined planes, vertical chutes,
circular cells. Finally, the success of the simple type of stability
analysis carried out in the present work calls for more accurate,
full-fledged approaches in which couplings of shear stress and
deformation with the density field would be taken into account.

\begin{acknowledgments}
  The authors thank M.\ Vennemann for his support and comments on the
  manuscript and L.\ Brendel for technical assistance and useful
  discussions. This work has been supported by DFG Grant No.\
  Wo577/8-1 within SPP 1486 ``Particles in Contact''.
\end{acknowledgments}



\end{document}